\documentclass[aps,superscriptaddress,groupedaddress,nofootinbib,12pt]{revtex4}  
\usepackage{hyperref}
\hypersetup{
    colorlinks=true,
    linkcolor=blue,
    filecolor=magenta,      
    urlcolor=cyan,
    citecolor=purple
    }
\usepackage{graphicx}  
\usepackage{dcolumn}   
\usepackage{bm}        
\usepackage{amssymb} \usepackage{amsmath}   
\usepackage[table]{xcolor}

\def\Mpl{M_{\textrm{Pl}}}

\begin{document}
\preprint{IPhT-t24/009}
\title{$\mathcal{R}^2$ effectively from Inflation to Dark
  Energy}
\author{Philippe~Brax and Pierre~Vanhove}
\address{CEA, DSM, Institut de Physique Th{\'e}orique, IPhT, CNRS, MPPU,
URA2306, Saclay, F-91191 Gif-sur-Yvette, France
}

\date{\today}

\begin{abstract}
 We consider the single-parameter $\mathcal{R}+ c\mathcal{R}^2$
 gravitational action and use constraints from astrophysics and the
 laboratory to derive a natural relation between the coefficient $c$
 and the value of the cosmological constant.  We find that the
 renormalisation of $c$ from the energy of the inflationary phase to
 the infrared, where the acceleration of the expansion of the Universe
 takes place, is correlated with the evolution of the vacuum
 energy. Our results suggest that the coefficient of the
 $\mathcal{R}^2$ term may provide an unexpected bridge between
 high-energy physics and cosmological phenomena such as inflation and
 dark energy.\\
 \textbf{This essay received an Honorable Mention in the 2024 Gravity Research Foundation Essay Competition.}
\end{abstract}

\maketitle
 
\section{Introduction}

Our understanding  of the force of gravity impacts  the way we  view  the
shape and the dynamics of the observable Universe. The current cosmological
paradigm relies on General Relativity, Einstein's theory of
gravity~\cite{Einstein:1916vd,Einstein:1915ca} crafted as a relativistic theory of
curved space-time. Gravity is inferred to be universal, i.e. it couples  equally to all types of matter and energy, from simple requirements such as Lorentz invariance~\cite{Weinberg:1964ew}.
Einstein proposed three classical tests of this theory of gravity~\cite{Will:2005va}:  the
perihelion precession of Mercury, the deflection of light by the Sun
and the gravitational redshift of light. These classical tests
illustrate the ubiquity of the gravitational force. In addition, several modern tests are routinely performed in order to test general relativity as well as effects that, in principle, could occur in a theory of gravitation different from Einstein's theory of gravity~\cite{ParticleDataGroup:2022pth}.

Einstein's General Relativity has stood the test of time and remains
unscathed after more than a century of intense investigations. In the
last 20 years, the emergence of the acceleration of the expansion of
the Universe~\cite{Copeland:2006wr,Clifton:2011jh,Weinberg:1988cp} has
led to various attempts to understand why gravity does not appear to
be attractive on large cosmological scales. This could lead to
intricate models of dark energy although the recent observation of the
neutron star merger by the LIGO/Virgo
consortium~\cite{LIGOScientific:2016lio,LIGOScientific:2017bnn} has
reinforced the strength of the claim that, so far, the best candidate
as an explanation for the cosmic acceleration is dark energy in the
form of a constant vacuum energy, whose archetype is the original
cosmological constant introduced by Einstein in 1917 and leading to
the de Sitter space-time of 1919.

In order the explain  the expansion of the  observable Universe a
cosmological constant is added  (with the signature $(-+++)$)
\begin{equation}
\label{e:EHLambda}
S_{EH-\Lambda}=\int d^4x \sqrt{-g} \left[ \frac{\Mpl^2}{2} \mathcal{R} -
  \Lambda^4 + g^{\mu\nu} T^{\rm matter}_{\mu\nu} \right] \, ,
\end{equation}
with a value today given by~\cite{ParticleDataGroup:2016lqr}
\begin{equation}\label{e:Lambda0}
\Lambda_0^4=\Omega_{\Lambda 0}\,3 ( H_0\Mpl )^{2}\simeq 2.4\times 10^{-47}
\textrm{GeV}^4\simeq\left(2.2 \times 10^{-3} \textrm{eV}\right)^4,
\end{equation}
where we have introduced $\Omega_{\Lambda 0}\sim 0.7$ as the fraction
of dark energy in the Universe and $\rho_c= 3\Mpl^2 H_0^2$ is the
critical density of the Universe. { Here $H_0$ is the Hubble rate
  today and $\Mpl=\sqrt{\hbar c/(8\pi G_N)}$ is the reduced Planck mass. }

The smallness of the cosmological constant is
still an open problem. This is  the ``old'' cosmological constant
problem~\cite{Weinberg:2000yb} whereby all the massive particles of
the Universe contribute to the vacuum energy with threshold
corrections which are quartic in their masses. Phase transitions, and
at least the electro-weak and the quantum chromodynamics ones, also
contribute to an alarming level.
These large contributions should  cancel in order to match with
observations for a determination of~\eqref{e:Lambda0} solely from the framework of particle physics.%

Concretely, the quantum corrections do read
\begin{equation}
\Lambda_0^4= \Lambda^4(m_e) -2\sum_{f=1}^3
\frac{m_f^4}{64\pi^2} \ln \frac{m_e^2}{m_f^2} + \delta \Lambda^4_0.
\label{e:rho-vac-me}
\end{equation}
where $\Lambda^4(m_e) $ encapsulates the contributions from
the tower of particle masses above the electron mass, i.e. it captures
the high energy physics effects which are relevant to the calculation
of the dark energy in the deep IR. We have singled out  the
contributions from the 
neutrino, because the neutrino masses are known to be at most in the milli-eV range
from the Planck 2018 constraints~\cite{Planck:2018vyg,Giusarma:2016phn,Vagnozzi:2017ovm,Giusarma:2018jei,DiValentino:2021imh,Esteban:2016qun}. We have added a component $\delta \Lambda_0^4$ coming from new physics at low energy below the electron mass. 

\noindent{\bf Cosmological constraints:}
The neutrino contribution to the vacuum energy
\begin{equation}
\delta\Lambda^4_{\rm neutrino}= -2\sum_{f=1}^3
\frac{m_f^4}{64\pi^2} \ln \frac{m_e^2}{m_f^2}.
\label{e:rho-vac-me2}
\end{equation}
can be estimated using the Planck data \cite{Planck:2018vyg} to be a large contribution
\begin{equation}\label{e:Lambdaneutrinorange}
  6\times 10^{-7} \textrm{eV}^4\leq |\delta \Lambda^4_{\rm neutrino}|\leq
  6\times 10^{-6}\textrm{eV}^4
\end{equation}
which exemplifies the nature of the cosmological constant problem even at low energy\footnote{ The Planck data constrains the sum of the neutrino masses $\sum_\nu m_\nu \le 0.12 {\rm eV}$ at 95 \% confidence level.}.

\noindent{\bf Astrophysical constraints:} We constrain the energy
densities $\Lambda^4(m_e)$ and $\delta \Lambda_0^4$ from astrophysical considerations.
The X-ray emitting gas of a galaxy cluster has
a typical temperature of $T_{\rm X} \sim 1$ keV, in regions of total
baryonic and dark matter density of about 500 times the mean density
of the Universe. These systems typically appeared at a redshift $z \gtrsim 0.1$
and already have a lifetime of the order of the age of the Universe.
In such clusters the neutrinos,  coming either from the early Universe with an energy of order $10^{-4}$ eV or from astrophysical processes such as the burning of stars
with an energy around $100$ keV, have a very small cross section with
matter and  decouple from the physics inside the clusters,
 which can then be described by  non-relativistic matter particles (such as electrons and protons)
 and General Relativity augmented with a vacuum energy. 
Under the strong assumption that the latter only  takes into account
all the physics for energy scales greater than $T_{\rm X}$, then the vacuum
energy $\vert\Lambda^4 (m_e)\vert$ cannot be too large otherwise it
would  significantly affect the dynamics within the cluster.
In a spherical approximation, the cluster would behave as a separate universe~\cite{Weinberg:1987dv}, with its own vacuum energy $\Lambda^4 (m_e)$.
To ensure small dynamical effects, we have the conservative bound
 \begin{equation}
  | \Lambda^4 (m_e) |   \lesssim 200 \,  \Lambda^4_0 .
   \label{lower}
 \end{equation}
{ This is the Weinberg bound which guarantees that the existence of galaxies is not prevented by the cosmological constant. Larger values would prevent the existence of galaxies and therefore clusters. The factor of 200 is obtained using the spherical collapse approximation for a separate Universe and reads more precisely~\cite{Weinberg:1987dv}
 \begin{equation}
  | \Lambda^4 (m_e) |   \lesssim \frac{500 }{729}\, \delta_R \rho_R.
   \label{lower2}
 \end{equation}
where $\rho_R$ is the matter density at recombination and $\delta_R$ the typical overdensity at the same epoch. }
This reasoning makes use of the presence at low redshift of hot high-density
structures within the cooler and lower-density cosmological background.

  \section{The $\mathcal{R}^2$ gravity}

The unknown contribution $\delta \Lambda^4_0$ is there to compensate for
the other quantum corrections to the vacuum energy in particular the large value from the neutrinos in~\eqref{e:rho-vac-me2}. The nature of
these extra quantum corrections have led to many possible scenarios
(supersymmetry, \dots) but so far have failed to give a convincing
answer.  In this text we propose a scenario from the purely
gravitational sector by embedding Einstein gravity into an effective
field theory framework.

It is very natural  to embed  the
classical Einstein theory of gravity into a quantum gravity framework valid at high energy, possibly close to the Planck scale, 
where the Einstein-Hilbert action~\eqref{e:EHLambda} is only the first term
of a low-energy effective action 
\begin{equation}\label{e:SeffGravity}
{\cal S}_{\rm eff} = \int d^4 x \sqrt{-g} \Bigg[\frac{\Mpl^2}{2} \mathcal{R} -
  \Lambda^4 + g^{\mu\nu} T^{\rm matter}_{\mu\nu} + \mathcal L_{\rm corrections}\Bigg]\,.
\end{equation}
The contributions $\mathcal L_{\rm corrections}$ arise from such  extensions
of Einstein gravity induced either by high-energy quantum corrections  from a high-energy completion like string theory
or new interactions from extra massless fields which are still undetected. { For instance in a string context and for curvatures below the string scale, only the curvature corrections up to cubic order are relevant. Higher order corrections in the curvature are suppressed by powers involving the ratio with the string scale~\cite{Burgess:2016owb,Brax:2017bcp}. Similarly the cubic term in the curvature is suppressed compared to the quadratic terms as long as the curvature is less than the compactification scale. In summary and as long as one considers scales below the string and compactification scales, models described by quadratic curvature corrections are sufficient. Of course this applies to the description of inflation as long as the inflation scale is low enough compared to the string and compactification scales.  }

This approach  is based on the idea that long range interactions in gravity can be described by a low-energy effective field theory, even if the high-energy behaviour of quantum gravity is still unknown~\cite{Donoghue:1995cz,Donoghue:2017pgk,Donoghue:2022eay}.
Although the status of  the  high-energy behaviour
of quantum gravity is still open, considering the effective field theory of gravity at
low energy does not pose a problem. One can safely extract low-energy
physics
from the quantization of the gravitational interactions observables
that are independent of the high-energy behaviour. 
We may quote J. D. Bjorken in~\cite{Bjorken:2000zz}  
who argues
that the Einstein-Hilbert term with its universal coupling to matter 
is naturally the first term of an effective field
theory of quantum gravity.  To the contrary to what many may think 
``as an open theory,
    quantum gravity is arguably our best quantum field theory, not the
    worst. [\dots] quantum
  gravity, when treated [\dots] as an effective field
  theory, has the largest bandwidth; it is credible over 60 orders of
  magnitude, from the cosmological to the Planck scale of
  distances.''~\cite{Bjorken:2000zz}.
This is precisely the philosophy that we will be following in this text.

Amongst the terms induced from the unknown high-energy
corrections, the scalar $\mathcal R^2$ operator plays a distinctive role
\begin{equation}
S_{R^2}(\mu)=\int d^4x \sqrt{-g} \left[ \frac{\Mpl^2}{2} \mathcal{R}
  -\Lambda^4(\mu)+ c_0(\mu)\mathcal{R}^2 +c_2(\mu)
\left(R_{\mu\nu}R^{\mu\nu}-\frac13 \mathcal{R}^2\right) \right]+ S_{\rm matter} \, ,
\end{equation}
it has
been shown in~\cite{Julve:1978xn,Fradkin:1981iu,Avramidi:1985ki,Salvio:2014soa} that the
coefficient $c_2(\mu)$ is always asymptotically free, since
$dc_2(\mu)/d \log \mu^2 > 0$, whereas $c_0(\mu)$ is asymptotically
safe, $dc_0(\mu)/d \log \mu^2 < 0$ for the non tachyonic case
$c_0(\mu) > 0$, which is the case of interest in this
paper. Therefore, at low-energy $c_2(\mu)$ tends to zero whereas
$c_0(\mu)$ grows, leading to the hierarchy $c_0(\mu) \gg c_2(\mu)$ at very low energy. Hence the quadratic Ricci scalar term is enhanced as compared with other quadratic and higher order contributions.

Therefore we can restrict ourselves to the scalar $\mathcal{R}^2$
\begin{equation}\label{e:SR2bis}
S_{R^2}(\mu)=\int d^4x \sqrt{-g} \left[ \frac{\Mpl^2}{2} \mathcal{R} -\Lambda^4(\mu)+ c_0(\mu) \mathcal{R}^2 \right] + S_{\rm matter} \, .
\end{equation}
At late time, a relevant effective action  in the same Wilsonian sense  is
\begin{equation}\label{e:Slate}
{\cal S}_{\rm late} = \int d^4 x \sqrt{-g} \Bigg[\frac{\Mpl^2}{2} \mathcal{R} -
  \Lambda_0^4+ c_0(\mu_{\rm late}) \, {\mathcal R}^2\Bigg]\,.
\end{equation}
Stelle~\cite{Stelle:1977ry} showed that the presence of the $\mathcal{R}^2$ term leads to
the propagation of a scalar massive degree of freedom, the scalaron, of mass
\begin{equation}
  m_{\rm scalaron}^2= {\Mpl^2\over 6c_0(\mu_{\rm late})},
\end{equation}
which could  contribute to a fifth force in gravitational experiments
  \begin{equation}\label{e:Vpot}
    V(r) = -{G_N M\over r}\, \left( 1+{1\over3}e^{-m_{\rm scalaron} r}\right)    \,.
  \end{equation}
  The absence of evidence for  short range forces  in the
E\"ot-Wash experiment~\cite{Hoyle:2004cw,Kapner:2006si,Lee:2020zjt} provides an upper bound on the range of scalar forces $d\le 52\mu{\rm m}$ corresponding to the strong lower
bound
\begin{equation}
m_{\rm scalaron}\gtrsim 3.8\times 10^{-3}~{\rm eV}\,.
\end{equation}
At this point we notice the observational coincidence that 
this lower bound  in the milli-eV range is strikingly close to the typical
energy scale set by the cosmological constant~\eqref{e:Lambda0}  and
suggest that the scalaron could lead  to the $\delta \Lambda_0^4$ contribution
to the vacuum energy~\eqref{e:rho-vac-me}. 
Or equivalently that the value of the coefficient of the $\mathcal{R}^2$ term could be
connected to the value of the cosmological constant.
The scalaron  contributes  to the vacuum energy as
\begin{equation}
  \delta \Lambda^4_{\rm scalaron}= {m^4_{\rm scalaron}\over 64\pi^2}
  \log\left(m_e^2\over m^2_{\rm scalaron}\right)  .
\end{equation}
Assuming that only the scalaron and the neutrinos have a mass smaller
than the electron mass, then $\delta \Lambda_0^4=\delta\Lambda^4_{\rm scalaron}$.
Combining the condition~\eqref{lower} with the
range~\eqref{e:Lambdaneutrinorange} we deduce that the scalaron mass
is bounded from above by the fourth-power mean value of the neutrino masses
\begin{equation}
m_{\rm scalaron}\lesssim \bar m_\nu=(m_1^4+m_2^4+m_3^4)^{1\over4}\simeq 0.1~\textrm{eV},
 \label{m-mnu-upper}
\end{equation}
leading to the narrow interval on the scalaron mass
\begin{equation}
  \label{e:massscalaroninterval}
  3.8\times 10^{-3} {\rm eV}\lesssim m_{\rm scalaron}
\lesssim 0.1\ {\rm eV}
\end{equation}
or for the coefficient  the $\mathcal{R}^2$ term in the gravity effective
action in the infrared (IR)
\begin{equation}
  \label{e:c0IRbound}
  10^{57}\lesssim c_0(\mu_{\rm IR}) \lesssim 7\times 10^{59}.
\end{equation}

\section{$\mathcal{R}^2$ running  from inflation to Dark Energy}

The $\mathcal{R}^2$ term plays an interesting role in two regimes of interest
for cosmology.
Its role could be significant from the large energies and early times of the inflationary era to the late time and low energy regime where dark energy sets the scene. 
Inflation is one of the most successful and elegant theoretical
descriptions of the post-Planck early time, since it overcomes in a
rigid way most of the shortcomings of the standard Big Bang
cosmology.   During
the inflationary era, the currently best model when compared to data is well described by the effective action with the $\mathcal{R}^2$ correction~\cite{Starobinsky:1980te}

\begin{equation}\label{e:Sinfl}
  \mathcal S_{\rm infl.} = \int d^4x \sqrt{-g}\left( {\Mpl^2\over2}
    \mathcal{R}  + c_0(\mu_{\rm infl})\mathcal{R}^2\right).
\end{equation}
This so-called   Starobinski's model produces a
  very good fit of the observed spectrum of primordial perturbations.
As the coefficient of the $\mathcal{R}^2$ term is dimensionless, it naturally 
depends on the energy scale in a Wilsonian sense, here set to be $\mu_{\rm infl}$ which can be identified with the nearly-constant Hubble rate during inflation.

Let us now review how to relate the inflation models~\eqref{e:Sinfl}  and
the late time effective action~\eqref{e:Slate} by the renormalisation
group equation.

The effective action in~\eqref{e:SR2bis} is an $f(\mathcal{R})$ theory
with
\begin{equation}
  \label{e:Fr}
  f(\mathcal{R})=-{2\Lambda^4(\mu)\over\Mpl^2}+\mathcal{R} + {2c_0(\mu)\over\Mpl^2} \mathcal{R}^2.
\end{equation}
This $f(\mathcal{R})$ model can be written as scalar-tensor theories with an action
\begin{equation}\label{e:Sscalaron}
  S_{\rm scalaron}= \int d^4x \sqrt{-g}\,\Mpl^2 \left(\frac12 \mathcal{R} -\frac{1}{2}(\partial \varphi)^2 - V(\varphi)\right) +S_{\rm matter}(e^{{2\over\sqrt6}\varphi}g_{\mu\nu})
\end{equation} 
where $\varphi$ is dimension less and the coupling to matter is  via the Jordan metric $g^J_{\mu\nu}=e^{{2\over\sqrt6}\varphi}g_{\mu\nu}$.
The potential in the action is given by the relation
\begin{equation}
V(\varphi)= \frac{\mathcal{R} \frac{df(\mathcal{R} )}{d\mathcal{R}}- f(\mathcal{R})}{( \frac{df(\mathcal{R} )}{d\mathcal{R}})^2}
\end{equation}
whilst the field $\varphi$ is related to the curvature as
\begin{equation}
  \frac{df(\mathcal{R} )}{d\mathcal{R}}=\exp\left(-{2\over \sqrt6}\varphi\right).
\end{equation}
The potential for the scalaron reads explicitly
\begin{equation}\label{e:Vscalaron}
V(\varphi)= \frac{\Mpl^2}{8c_0(\mu)}\left(e^{{2\varphi\over\sqrt6}} -1\right)^2 +2 \frac{\Lambda^4(\mu)}{\Mpl^2} e^{{4 \varphi\over\sqrt6}}
\end{equation}
as a function of the scalaron $\varphi$. 
From this potential we can read-off the effective mass of the scalaron
\begin{equation}\label{e:massscalaron}
    m^2_{\rm scalaron}(\mu)= {\Mpl^2\over 6c_0(\mu)}+{16\over
      3}{\Lambda^4(\mu)\over \Mpl^2}.
\end{equation}
Having fixed the Planck mass $\Mpl^2$, which is then
independent of $\mu$ and fixes all the energy, distance and time  scales, the parameters in the
Wilson effective action~\eqref{e:Sscalaron} are $\mu$ dependent.  The
cosmological constant $\Lambda(\mu)$ and the coefficient $c_0(\mu)$ of the
$\mathcal{R}^2$ term follow the renormalisation group equations
where they evolve each time a particle species of mass $m$ is
integrated out~\cite{Georgi:1993mps}, i.e. when $\mu \le m$. In the
history of the Universe, particles in the thermal bath are integrated
out when the temperature falls below the mass $m$, leading to
\begin{equation}
\frac{d   m^2_{\rm scalaron}(\mu)}{d\ln \mu}= -\frac{1  }{48 \pi^2 \Mpl^2}{\cal S}tr( M^4) \theta(\mu-M).
\end{equation}
and
\begin{equation}
\frac{d\Lambda^4(\mu)}{d\ln \mu}= -\frac{ 1 }{32 \pi^2}{\cal S}tr( M^4)\theta(\mu-M)
\end{equation}
in terms of the supertrace of the complete mass matrix $M$ of all the particles in the Universe. 
The scalaron arising from the gravitational sector of the theory
couples to all particles in the spectrum.
This implies that the coefficient of the $\mathcal{R}^2$ term has an
evolution linked to the one of the cosmological constant
\begin{equation}\label{e:RGeq}
\frac{ d (\Mpl^4  c_0(\mu)^{-1}+28\Lambda^4(\mu))}{d\ln \mu}=0\,.
\end{equation}
The initial
values of the renormalisation group are taken at the end of
inflation corresponding to the reheat temperature $T_{\rm reh}$, i.e. larger than any physical masses of the particles in the
spectrum of the theory.  { The previous renormalisation group equations allow one to calculate the evolution of $c_0$ and $\Lambda$ from inflation to the infrared. The only new ingredient in this integration is the presence of phase transitions taking place at different values of $\mu$, for instance the QCD phase transition, where the cosmological constant jumps due to the change of vacuum energy during the transitions~\cite{Bellazzini:2015wva}. }

The value of $c_0(\mu_{\rm end})$ at the end of inflation can be deduced from the normalisation of the CMB spectrum as
\begin{equation}
 \frac{V(\varphi_\star)^3}{\Mpl^2(V'(\varphi_\star))^2}\simeq
 {3e^{-{4\varphi_\star\over \sqrt6}}\over32
   c_0(\mu_{\rm end})}\simeq 2\times 10^{-11}
\end{equation}
evaluated at the value of $\varphi_\star$ determined by   the spectral index
\begin{equation}\label{e:ns}
n_s-1= 2 \eta_\star  =2
\frac{V''(\varphi)}{V(\varphi)}\Big|_{\varphi=\varphi_\star} \simeq
-{8\over 3} e^{2\varphi_\star\over\sqrt6}.
\end{equation}
giving the constraint on  the coefficient of the $\mathcal{R}^2$ from the CMB data
\begin{equation}\label{cinf}
  c_0^{-1}(\mu_{\rm end})\simeq 1.52\, 10^{-9}\, (n_s-1)^2\simeq
2\times 10^{-12},
\end{equation}
where we used that $n_s-1\simeq -0.0351$ according to the Planck
data~\cite{Planck:2018vyg}.

 The cosmological constant $\Lambda^4(\mu_{\rm end})$ at the end of inflation
 is not directly determined by the experimental data, but from an ultraviolet (UV)
 point of view, its value is constrained by the renormalisation group equation~\eqref{e:RGeq}  reinforcing the fact that the physics at high energy
seems to be largely constrained by the physics at low energy. This is
the case of the coefficient of the $\mathcal{R}^2$ term during inflation which is constrained by the CMB data. Here  the physics of the vacuum in the IR, i.e.\ the vacuum stability
combined with gravitational tests, determines indirectly the value of
the cosmological constant in the UV. Of course, our analysis does not
provide any explanation for this value from a top-bottom point of
view.

\section{Observational predictions}
\label{sec:pred-scal-mass}

The gravitational effective action~\eqref{e:SR2bis} displays an
interesting universality behaviour as it can be applied to the inflation
regime and the late cosmological times.
The effective action has a priori two free parameters the cosmological
constant $\Lambda^4(\mu)$ and the coefficient $c_0(\mu)$ of the $\mathcal{R}^2$
term. We have argued that these two coefficients have an evolution
that is related by the renormalisation group equation
in~\eqref{e:RGeq} and determined the value of $c_0(\mu)$ 
in the UV and IR from observations.

At late time, the coefficient of the $\mathcal{R}^2$ term is in the narrow
interval for the scalaron mass~\eqref{e:massscalaroninterval}.
This prediction can be tested by the E\"ot-Wash experiments~\cite{Hoyle:2004cw,Kapner:2006si,Lee:2020zjt}.

We remark that  this interval is compatible with the
value $m_{\rm scalaron}\simeq 4.4\times 10^{-3}$ eV for which the
$\mathcal{R}^2$ induced scalaron could
be at the origin of the observed dark matter
abundance~\cite{Cembranos:2008gj,Cembranos:2015svp,Shtanov:2021uif,Shtanov:2022xew}.
In this scenario,  at low energy compared to the inflation scale, the vacuum
expectation value of the scalaron is displaced from the
origin by an amount depending on the electroweak scale $v\simeq 250$~GeV because of its coupling to the Higgs field.
For more generic initial conditions after inflation taking into
account the quantum fluctuations of the scalaron during inflation, the
whole interval up to $m_{\rm scalaron}\simeq 0.1$ eV could accommodate
dark matter. As the electroweak transition begins, the scalaron starts
oscillating with a decreasing amplitude eventually converging to the
origin. This misalignment mechanism is similar to what happens for
axions and leads an abundance of dark matter which fits the observed
value for $m_{\rm scalaron} \simeq 4.4\times 10^{-3}$~eV~\cite{Shtanov:2022xew}. Combining both scenarios, this would lead
to a possible signal in gravitational experiments below a distance
$d\lesssim 45~\mu{\rm m}$, which is 
well within the allowed interval~\eqref{e:massscalaroninterval}
\footnote{ The link with scalar dark matter is guaranteed as long as the amplitude of oscillation around the minimum of the scalar potential is small and higher order terms act as corrections to a simple quadratic oscillator. When the quadratic terms are dominant the scalar dark matter is of the fuzzy type~\cite{Hui:2016ltb} whilst corrections to the quadratic Lagrangian can also be interesting astrophysically~\cite{Brax:2019fzb}. We leave a detailled study of the phenomenology for future work.} . This strongly supports the exciting possibility of testing the existence of the new gravitational interaction ${\cal R}^2$, which could play a role in both the late acceleration of the universe and dark matter. All in all, we believe that this a  is a consequence of the
universality of the effective long distance physics of gravity compared to the scales of quantum gravity. 

\end{document}